\documentclass[lettersize,journal]{IEEEtran}
\usepackage{amsmath,amsfonts,amsfonts, subfigure}
\usepackage{cite}
\usepackage{algorithmic}
\usepackage{graphicx}
\usepackage{textcomp}
\usepackage{epstopdf}
\usepackage{array}
\usepackage{stfloats}
\usepackage{multirow, booktabs, bbding, url, pifont}
\usepackage{verbatim}
\usepackage{balance}
\usepackage{xcolor}
\def\BibTeX{{\rm B\kern-.05em{\sc i\kern-.025em b}\kern-.08em
    T\kern-.1667em\lower.7ex\hbox{E}\kern-.125emX}}

\begin{document}
\title{Learning Lossless Compression for \\High Bit-Depth Volumetric Medical Image}
\author{Kai Wang, Yuanchao Bai, \IEEEmembership{Member, IEEE}, Daxin Li, Deming Zhai, \\Junjun Jiang, \IEEEmembership{Senior, IEEE}, Xianming Liu, \IEEEmembership{Member, IEEE}
\thanks{
The authors are with Faculty of Computing, Harbin Institute of Technology, Harbin, 150001, China (e-mail: cswangkai@stu.hit.edu.cn; yuanchao.bai@hit.edu.cn; hahalidaxin@stu.hit.edu.cn; \{zhaideming, jiangjunjun, csxm\}@hit.edu.cn). Xianming Liu is the corresponding author.}
}
\markboth{Journal of \LaTeX\ Class Files,~Vol.~XX, No.~XX, September~2024}%
{Learning Lossless Compression for \\High Bit-Depth Volumetric Medical Image}

\maketitle
\begin{abstract}
Recent advances in learning-based methods have markedly enhanced the capabilities of image compression. However, these methods struggle with high bit-depth volumetric medical images, facing issues such as degraded performance, increased memory demand, and reduced processing speed. To address these challenges, this paper presents the Bit-Division based Lossless Volumetric Image Compression (BD-LVIC) framework, which is tailored for high bit-depth medical volume compression. The BD-LVIC framework skillfully divides the high bit-depth volume into two lower bit-depth segments: the Most Significant Bit-Volume (MSBV) and the Least Significant Bit-Volume (LSBV).
The MSBV concentrates on the most significant bits of the volumetric medical image, capturing vital structural details in a compact manner. This reduction in complexity greatly improves compression efficiency using traditional codecs. Conversely, the LSBV deals with the least significant bits, which encapsulate intricate texture details. To compress this detailed information effectively, we introduce an effective learning-based compression model equipped with a Transformer-Based Feature Alignment Module, which exploits both intra-slice and inter-slice redundancies to accurately align features. Subsequently, a Parallel Autoregressive Coding Module merges these features to precisely estimate the probability distribution of the least significant bit-planes. Our extensive testing demonstrates that the BD-LVIC framework not only sets new performance benchmarks across various datasets but also maintains a competitive coding speed, highlighting its significant potential and practical utility in the realm of volumetric medical image compression.

\end{abstract}


\section{Introduction}
\label{sec: introduction}
\IEEEPARstart{T}{he} ever-increasing utilization of advanced medical imaging techniques, such as computer tomography (CT) scans and magnetic resonance imaging (MRI), has led to a surge in the generation of volumetric medical image data. 
For instance, a single CT scan can produce a three-dimensional (3D) image of a patient's body comprising dozens or even hundreds of two-dimensional (2D) slices at a 16-bit depth, resulting in a total data volume reaching hundreds of megabytes or even one gigabyte. These detailed volumetric images are indispensable for accurate diagnosis and effective treatment planning. However, their considerable size presents notable challenges for storage and transmission.
Given these challenges, there is a clear need for efficient compression methods. While lossy compression techniques can achieve higher compression ratios, they risk introducing distortions that could compromise the diagnostic integrity of the images and potentially lead to medical errors. Consequently, lossless compression emerges as the preferred method for medical imaging, adhering to the strict requirements for maintaining data fidelity.

A variety of techniques for the lossless compression of volumetric medical images have been explored, broadly categorized into 2D-based and 3D-based codecs. 2D-based codecs \cite{weinberger2000loco,calic, sneyers2016flif, alakuijala2019jpeg, skodras2001j2k}, focus on minimizing redundancy within individual image slices. In contrast, 3D-based codecs aim to leverage the shared information between adjacent slices to achieve further compression.
Among 3D codecs, there are principally two techniques used to exploit the correlation between slices: wavelet-based and video-based codecs. Wavelet-based methods, exemplified by JP3D \cite{schelkens2006jpeg2000p10}, employ the 3D DWT to leverage redundancy both within and across slices. On the other hand, video-based codecs, such as those employing High Efficiency Video Coding (HEVC) \cite{parikh2017highHEVC}, use motion field estimation to capitalize on the inter-slice redundancy. However, both strategies face challenges due to the complex structural variations present in consecutive slices, which complicates both 3D transforms and motion estimation processes. Moreover, traditional codecs depend on manually designed modules that are not amenable to end-to-end optimization, thereby constraining their compression efficiency. This limitation highlights the need for effective approaches that can dynamically adapt to the complex characteristics of volumetric medical data.

Recent research has explored learning-based methods for lossless image compression of natural images. These methods leverage deep generative models \cite{pixelrnn2016icml,pixelcnn_pp,zhang2021out,rhee2022lcfdnet,idf2019nips,zhang2021nipsiflow,iclr2019bitback,bai2021learning,bai2022deep,dardouri2021dynamic} to estimate the probability distribution for each pixel. Subsequently, entropy coding techniques, such as Arithmetic Coding (AC) \cite{arithmetic_coding} and Asymmetric Numeral Systems (ANS) \cite{duda2009asymmetric}, utilize these probability distributions to convert images into bitstreams. The length of each probability distribution typically aligns with the numerical range of the corresponding pixel.
Unlike these learned lossless 2D compression methods that solely target intra-slice redundancy, learned volumetric image compression techniques \cite{lanz2019graph,nagoor2020lossless,chen2022exploitingicec,xue2022aiwave,liu2024bilateral} additionally exploit inter-slice information to enhance their generative models. 
However, in high bit-depth medical images, each pixel's expansive numerical range (often reaching $2^{16}$) necessitates extensive probability distributions. \textcolor{black}{Consequently, applying these existing methods directly results in larger distribution tables and demands for high-precision entropy coding. This leads to performance degradation, increased memory requirements, and significantly lower compression throughputs.}

\begin{figure}[!t]
\centering
\includegraphics[width=0.9\linewidth]{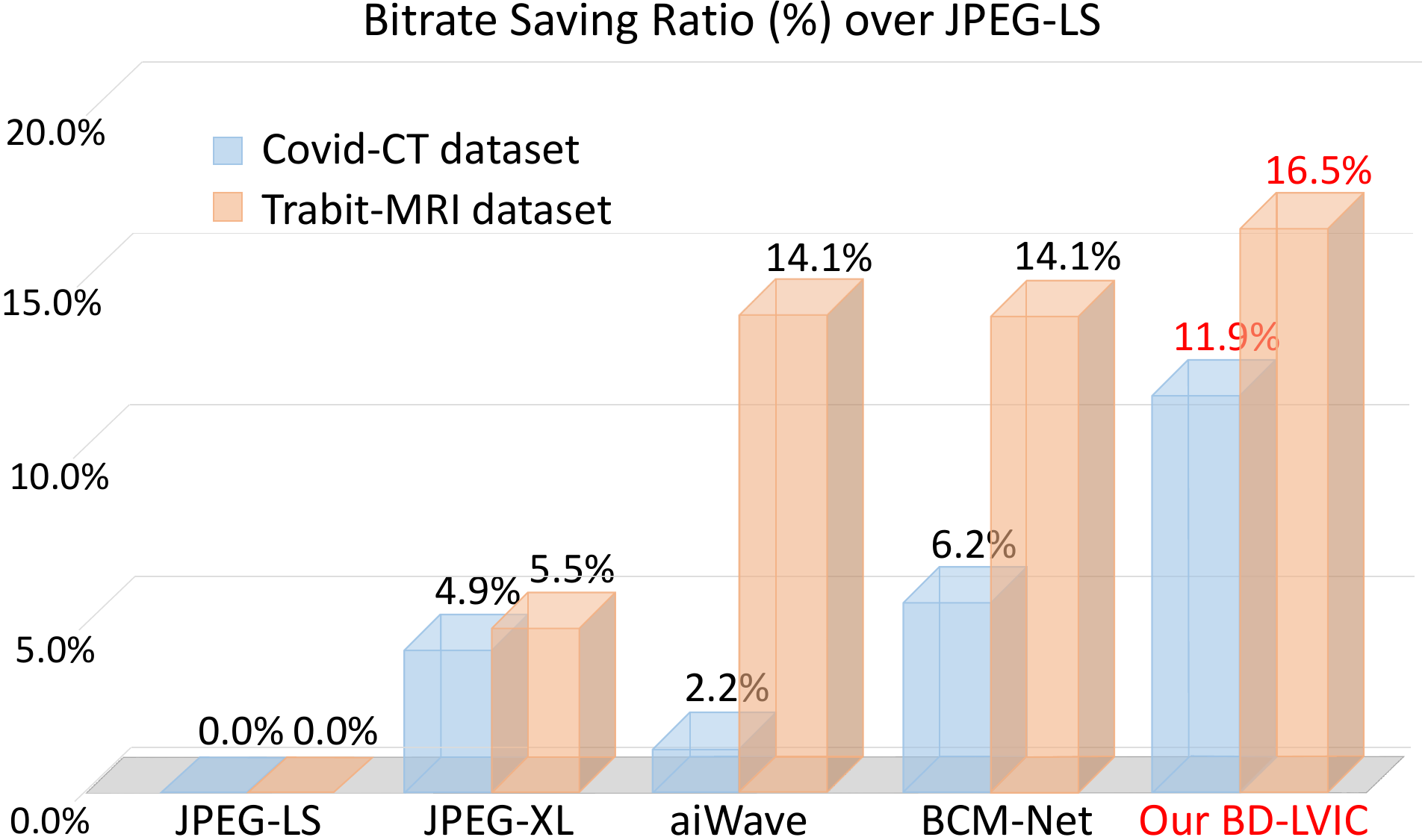}
\setlength{\abovecaptionskip}{-1pt}
\caption{
Comparative analysis of the bitrate saving ratios on the Covid-CT \cite{morozov2020mosmeddatacovid} and Trabit-MRI \cite{mader2019trabit2019} datasets, using the JPEG-LS \cite{weinberger2000loco} codec as the baseline for calculation. Among the compared codecs, JPEG-XL \cite{alakuijala2019jpeg} stands as the most advanced traditional image codec, in contrast to aiWave \cite{xue2022aiwave} and BCM-Net \cite{liu2024bilateral}, which represent the cutting-edge in learned lossless compression methods for volumetric medical images.
}
\label{fig: bitrate_saving_ratio}
\vspace{-0.2cm}
\end{figure}

In this study, we address the outlined challenges by introducing the  Bit-Division Based Lossless Volumetric Image Compression (BD-LVIC) framework, specifically designed for high bit-depth volumetric medical images. Our framework ingeniously partitions a high bit-depth medical volume into two lower bit-depth subvolumes, thus narrowing the numerical range: the Most Significant Bit-Volume (MSBV) and the Least Significant Bit-Volume (LSBV).
The MSBV contains the most significant bit-planes of the medical volume, encapsulating crucial structural information in a sparse format that is highly amenable to compression. For this component, we employ conventional codecs, capitalizing on their low computational complexity for efficient compression.
Conversely, the LSBV, which is primarily composed of complex textures, poses more substantial compression challenges. To tackle this, we develop a novel learned compression model that processes each Least Significant Bit-Slice (LSBS) within the LSBV on a slice-by-slice basis.
For the encoding of each LSBS, we design a Transformer-Based Feature Alignment Module (TFAM), which generates aligned features by integrating information from the corresponding Most Significant Bit-Slice (MSBS), as well as the preceding MSBS and LSBS, leveraging both intra-slice and inter-slice redundancies through sophisticated cross-attention and self-attention mechanisms.
Following feature alignment, a Parallel Autoregressive Coding Module (PACM) merges these aligned features with the local spatial context, obtained through masked convolution, to accurately estimate the probability distribution of the current LSBS. This facilitates efficient entropy coding. Finally, the compressed representation of the volumetric medical image comprises the aggregated bitstreams of both MSBV and LSBV. 
As illustrated in Fig.\;\ref{fig: bitrate_saving_ratio}, the innovative design of our BD-LVIC framework ensures outstanding compression efficiency on a range of volumetric medical datasets.


The main contributions are summarized as follows:
\begin{itemize}

\item We propose a novel BD-LVIC framework for high bit-depth volumetric medical image compression that strategically decomposes a high bit-depth volume into two lower bit-depth subvolumes \textcolor{black}{and specifically designs the joint compression of the two subvolumes. The two subvolumes are represented as the Most Significant Bit-Volume and the Least Significant Bit-Volume.}

\item For efficient LSBV compression, we propose a learned model with two key components: TFAM and PACM. The TFAM leverages both intra-slice and inter-slice dependencies to generate aligned features. The PACM then extracts local spatial context and fuses it with these features, enabling the model to effectively model the probability distribution of the current LSBS in a parallel scan order.

\item Extensive experiments on various medical image datasets demonstrate that our BD-LVIC framework achieves state-of-the-art lossless compression performance and fast coding speed, underscoring its considerable promise and practical value for medical imaging applications.

\end{itemize}

This paper presents a significant and non-trivial advancement over our prior work \cite{wang2023learning}, marking a pivotal evolution in our research trajectory. In \cite{wang2023learning}, our focus was on the compression of 2D medical images. This paper extends our exploration into the realm of volumetric medical image compression, addressing the unique challenges and leveraging the opportunities presented by three-dimensional data. A cornerstone of this advancement is the introduction of a novel learned compression model that effectively utilizes both intra-slice and inter-slice redundancies. Moreover, diverging from sequential coding in raster scan order \cite{wang2023learning}, we unveil the analytical relationship between parallel coding and scanning angle, and introduce a parallel autoregressive coding scheme that significantly expedites the LSBV coding progress.

The remainder of this paper is organized as follows: Section \uppercase\expandafter{\romannumeral2} reviews related works. We introduce observation and motivation  in Section \uppercase\expandafter{\romannumeral3}. The proposed BD-LVIC framework is detailed in Section \uppercase\expandafter{\romannumeral4}. Experimental results and the conclusion are presented in Sections \uppercase\expandafter{\romannumeral5} and \uppercase\expandafter{\romannumeral6}, respectively.

\section{Related Work}
\label{sec: related work}
\subsection{Traditional Lossless Codecs}
\subsubsection{2D Image Codecs}
Traditional 2D image codecs are designed to reduce intra-slice redundancy \cite{weinberger2000loco,calic, sneyers2016flif, alakuijala2019jpeg, skodras2001j2k}. JPEG-LS \cite{weinberger2000loco} implements the pixel-based prediction method, which predicts each pixel based on its neighbors and encodes the residual effectively. JPEG2000 \cite{skodras2001j2k} applies an invertible discrete wavelet transformation that decomposes images into different frequency bands, thereby increasing the similarity within these bands and enhancing compression. Moreover, JPEG-XL \cite{alakuijala2019jpeg} leverages global sampling strategies for constructing decision trees that provide adaptive weights for each pixel. This approach significantly improves prediction accuracy and achieves state-of-the-art performance in conventional image codecs.

\subsubsection{Volumetric Image Codecs}
Traditional 3D volumetric image codecs \cite{jp2kp2, schelkens2006jpeg2000p10, bruylants2015wavelet3ddwt,starosolski2020employing3ddwt, niedermayer2013ffv1, parikh2017highHEVC, lucas2017lossless3dpredictors} enhance compression efficiency by exploiting inter-slice redundancy. Prominently, 3D-DWT-based codecs \cite{schelkens2006jpeg2000p10, bruylants2015wavelet3ddwt,starosolski2020employing3ddwt} are used for compressing volumetric medical images, leveraging 3D-DWT transformations to address both intra-slice and inter-slice redundancy. Lucas \emph{et al.} \cite{lucas2017lossless3dpredictors} proposed innovative 3D predictors for this purpose. Sanchez \emph{et al.} \cite{sanchez2009symmetrytmi} employed symmetry-based techniques to minimize redundancy in volumetric medical imaging.
Moreover, video-based codecs like High Efficiency Video Coding (HEVC) \cite{parikh2017highHEVC} and Versatile Video Coding (VVC) \cite{bross2021overviewvvc} apply motion field estimation between slices to reduce inter-slice redundancy. 
Nevertheless, the compression capabilities of traditional codecs are somewhat constrained due to their reliance on manually-crafted modules, which cannot be optimized in an integrated end-to-end fashion.
\textcolor{black}{Currently, there are also 3D compressors designed for immersive video \cite{boyce2021mpegtmiv} and point clouds \cite{graziosi2020overviewgpccvpcc}. However, the input requirements of multi-view and depth maps for immersive videos, or the inherent sparsity of point clouds, render these 3D compressors unsuitable for compressing volumetric medical images.}

\begin{figure*}[!t]
\centering
\subfigure [PMF construction time vs. bit depth]{
\includegraphics[width=0.31\linewidth]{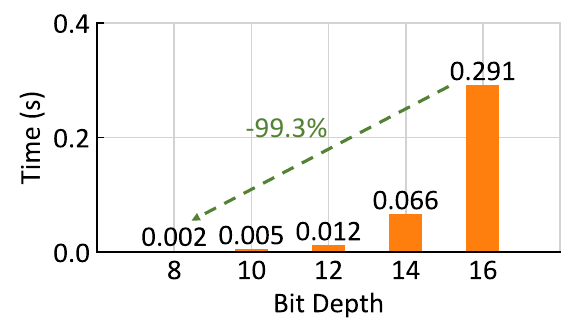}
\label{fig:motivation:sub1}
}
\hspace{0.3mm}
\subfigure [Memory usage vs. bit depth]{
\includegraphics[width=0.31\linewidth]{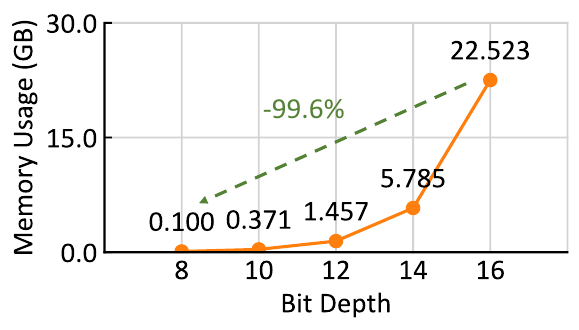}
\label{fig:motivation:sub2}
}
\hspace{0.3mm}
\subfigure [PMF construction time vs. patch size, bit depth]{
\includegraphics[width=0.31\linewidth]{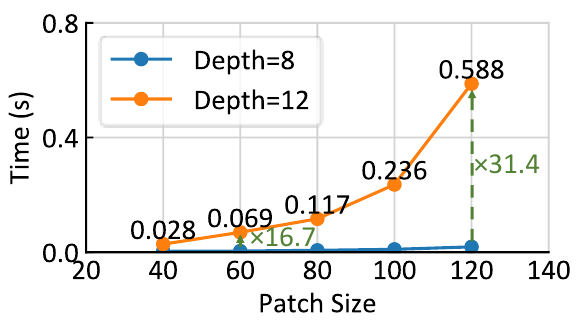}
\label{fig:motivation:sub3}
}
\caption{
(a) Assessing the impact of bit depth on PMF construction time, (b) Exploring bit depth's effect on memory usage, (c) Investigating the impact of patch size and bit depth on PMF construction time.}
\label{fig:motivation}
\vspace{-0.2cm}
\end{figure*}

\subsection{Learning-based Lossless Codecs}
The rapid progress in deep learning has catalyzed the development of various learning-based lossless codecs for both natural and volumetric images.

\subsubsection{Learned 2D Image Codecs}
In the field of learned lossless codecs for natural images, a variety of deep generative models are utilized, including autoregressive models \cite{pixelrnn2016icml,pixelcnn_pp,rhee2022lcfdnet}, flow-based models \cite{idf2019nips,zhang2021nipsiflow}, and variational auto-encoder models (VAE) \cite{iclr2019bitback,bai2021learning,bai2022deep}. Their critical function is in the precise estimation of image data's probability distribution. Following this estimation, entropy coding techniques, specifically Arithmetic Coding (AC) \cite{arithmetic_coding} and Asymmetric Numeral Systems (ANS) \cite{duda2009asymmetric}, are employed to efficiently encode images into compact bitstreams. Notably, these models designed for 8-bit natural images require specific adaptations to suit high bit-depth medical images.

\subsubsection{Learned Volumetric Image Codecs}

Recent advances in learned volumetric image codecs \cite{lanz2019graph,nagoor2020lossless,chen2022exploitingicec,xue2022aiwave,liu2024bilateral} have emphasized exploiting intra-slice and inter-slice redundancies to augment deep generative model performance. Specifically, Nagoor \emph{et al.} \cite{nagoor2020lossless} developed a 3D predictor with local sampling for volumetric data, but its compression efficiency is hindered by its network design. Chen \emph{et al.} presented ICEC \cite{chen2022exploitingicec}, which utilizes gating mechanisms to integrate multi-scale features. However, this method shows limited effectiveness in exploiting intra- and inter-slice data, impacting its overall performance. Xue \emph{et al.} proposed aiWave framework \cite{xue2022aiwave}, which applies 3D affine wavelet-like transformations to reduce volumetric data redundancies, but its lengthy coding time (average 900 seconds per slice) constrains its practical application. Liu \emph{et al.} adopted a lossy plus residual coding framework for volumetric data, using VVC \cite{bross2021overviewvvc} as the lossy component and introducing a bilateral context model for residual compression. However, the inability to jointly optimize both lossy and lossless components constrains the performance of their method.

\section{Observation and Motivation}
\label{sec: proposed method}

Through extensive data analysis, we observe that two key challenges arise when encoding high bit-depth volumetric medical images: 1) the inherent curse of high bit-depth, and 2) the combined effects of intra-slice and inter-slice information redundancy within volumetric data. These critical challenges motivate our research, driving the development of the BD-LVIC framework presented in this paper.

\subsubsection{Curse of High Bit Depth}
As bit depth increases, the size of Probability Mass Function (PMF) tables grows exponentially. This significantly hinders accurate probability distribution estimation, raising computational complexity and memory demands during encoding and decoding. Consequently, coding process suffers considerably, limiting the feasibility of implementing the method on edge devices with resource constraints.

We begin by evaluating the impact of varying bit depths on PMF construction time and GPU memory occupancy, while maintaining a consistent image resolution.
We crop single-channel images with a resolution of $30\times30$ and generate the parameters of the mixture logistic distribution. The bit depths are set to $\{8, 10, 12, 14, 16\}$. As depicted in Fig.\;\ref{fig:motivation:sub1}, the PMF construction time for a 16-bit image is $99.3\%$ higher compared to an 8-bit image, leading to a significant decrease in coding speed. Concurrently, as shown in Fig.\;\ref{fig:motivation:sub2}, memory consumption escalates exponentially with increasing bit depth. 
In addition, we investigate the impact of increasing image resolution (number of pixels) on PMF construction time for fixed bit depths (8-bit and 12-bit).
As shown in Fig.\;\ref{fig:motivation:sub3}, increasing the resolution from $60\times 60$ to $120\times 120$ pixels leads to a $16.7$-fold increase in PMF construction time for 8-bit images and a $31.4$-fold increase for 12-bit images.
It is important to note that typical CT images have a higher resolution, often reaching $512\times512$ pixels. For such high resolutions, PMF construction time for high bit-depth medical images would be significantly slower compared to 8-bit images.
This motivates the development of our proposed bit-division based image compression framework. This framework decomposes high bit-depth volumes into two lower bit-depth subvolumes: MSBV and LSBV.

\subsubsection{Intra- and Inter-slice Redundancy in Volume}
\label{sec: two redundancies}
\begin{figure}[!t]
\centering
\subfigure [$X_t$]{
\includegraphics[width=0.27\linewidth]{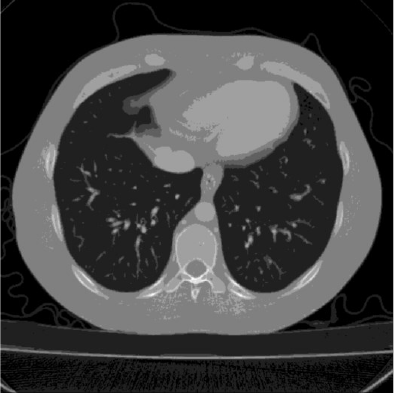}
\label{fig:visual_image:org}
}\hspace{-2mm}
\subfigure [$X_t^M$]{
\includegraphics[width=0.27\linewidth]{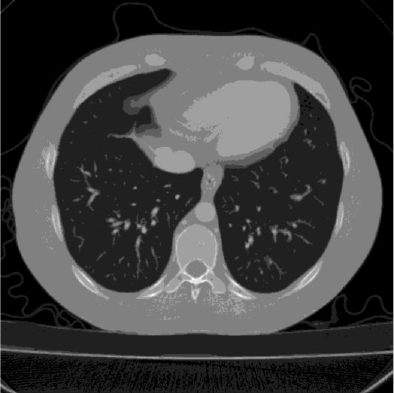}
\label{fig:visual_image:msb}
}\hspace{-2mm}
\subfigure [$X_t^L$]{
\includegraphics[width=0.27\linewidth]{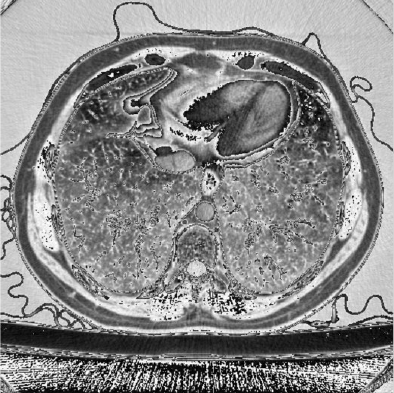}
\label{fig:visual_image:lsb}
}
\subfigure [$X_{t+1}$]{
\includegraphics[width=0.27\linewidth]{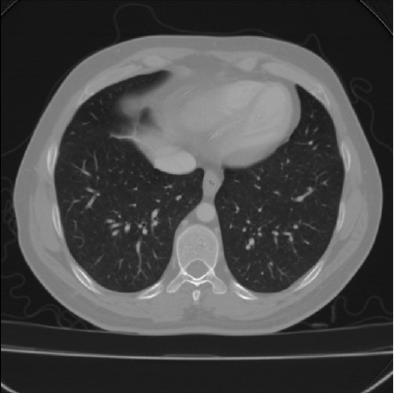}
}\hspace{-2mm}
\subfigure [$X_{t+2}$]{
\includegraphics[width=0.27\linewidth]{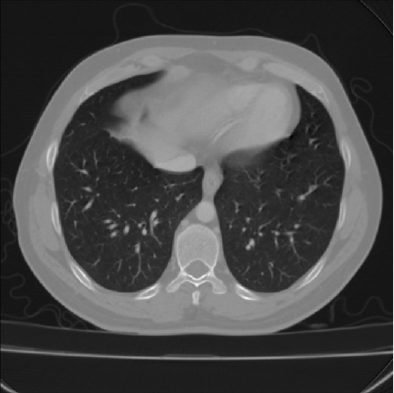}
}\hspace{-2mm}
\subfigure [$X_{t+2}-X_{t+1}$]{
\includegraphics[width=0.27\linewidth]{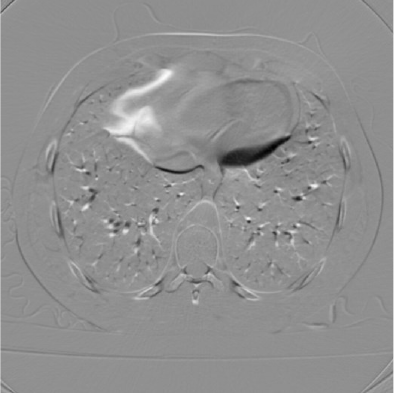}
}
\setlength{\belowcaptionskip}{10 pt}
\caption{(a-c) The high bit-depth medical slice of abdominal CT, denoted by $X_t$, alongside its corresponding MSBS $X_t^M$ and LSBS $X_t^L$. Here, $t$ denotes the index of the current slice. (d-f) Adjacent slices $X_{t+1}$, $X_{t+2}$, and their difference $X_{t+2}-X_{t+1}$.}
\label{fig:visual_image}
\vspace{-0.1cm}
\end{figure}

Fig.\;\ref{fig:visual_image}(a)-(c) demonstrate significant intra-slice redundancy between MSBS and LSBS within MSBV and LSBV, because of their shared structural characteristics. However, exploiting this correlation for compression is challenging  because MSBS and LSBS have distinct data characteristics and value ranges. Additionally, Fig.\ref{fig:visual_image}(d)-(f) highlight the substantial inter-slice redundancy between adjacent slices. Yet, unlike typical video sequences with motion-based deformations, the structural similarities observed in adjacent slices of medical images often arise from anatomical progressions through the body. This distinction presents a unique challenge for volumetric image compression.

To efficiently harness these two forms of intricate redundancy, we propose a novel learned compression model with two key components: TFAM and PACM. The TFAM leverages both intra-slice and inter-slice dependencies to generate aligned features. The PACM then extracts local spatial context and fuses it with these features, enabling the model to effectively model the probability distribution of the current LSBS in a parallel scan order.

\begin{figure*}[!t]
\centering
\includegraphics[width=0.99\linewidth]{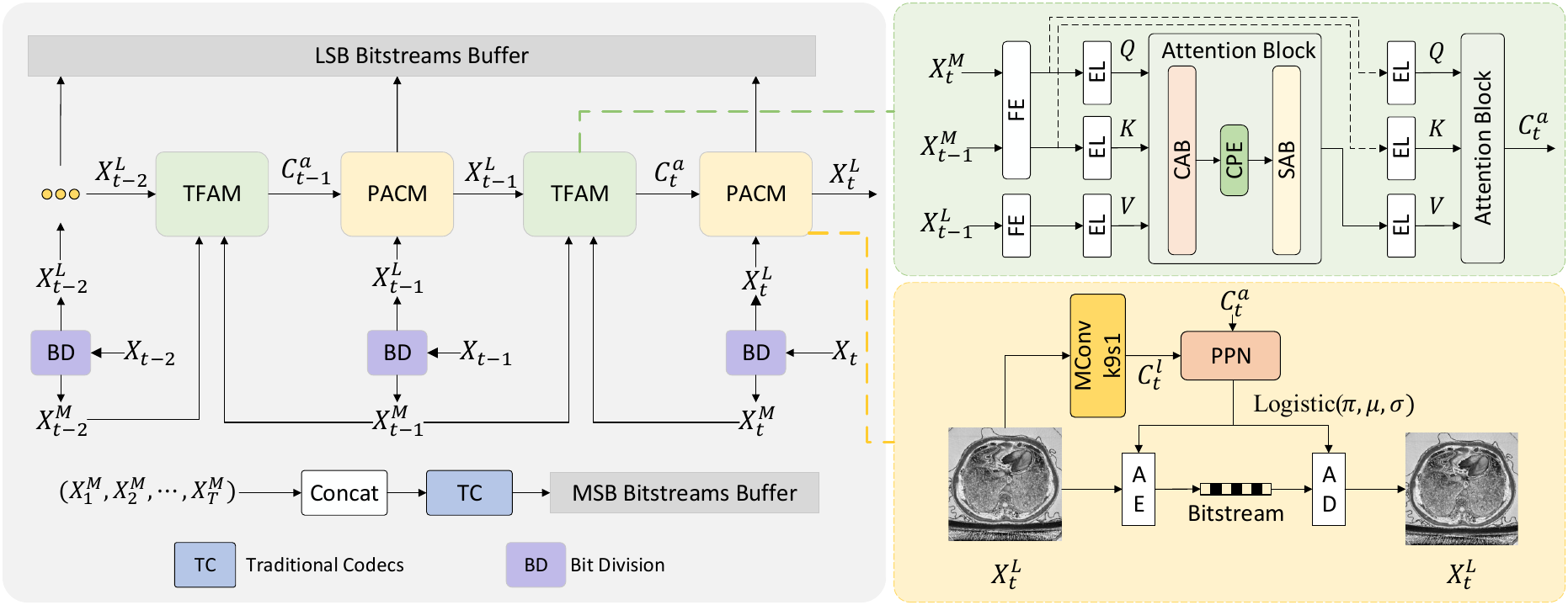}
\caption{The overview of our BD-LVIC framework. Each slice of medical volume is decomposed into MSBS and LSBS. We first utilize traditional codecs to compress all MSBSs. Then, each LSBS is encoded slice-by-slice. During encoding the current LSBS $X_t^L$, we employ TFAM to generate the aligned feature $C_t^a$ and utilize PACM to extract local context $C_t^l$ and fuse it with $C_t^a$ to estimate the distribution of $X_t^L$. 
TFAM includes Feature Extraction (FE), Embedding Layer (EL), Self-attention Block (SAB), Conditional Position Embedding (CPE), Cross-attention Block (CAB), and PACM contains Masked Convolution (MConv), Parameter Predictor Network (PPN), Arithmetic Encoder (AE), Decoder (AD).}
\label{fig: overview}
\vspace{-0.1cm}
\end{figure*}

\section{BD-LVIC Framework}
In this section, we first provide an overview of our BD-LVIC framework. Next, the compression process for the MSBV is described. Following this, we delve into the compression strategy for the LSBV, detailing the design and implementation of two crucial modules: TFAM and PACM.
\subsection{Overview}
Let $\mathbf{X}=(X_{1}, X_{2},\cdots, X_{T})$ represents the high bit-depth volume with $T$ slices, where $X_{t}$ represents the $t$-{th} slice of the volume. We decompose $\mathbf{X}$ into two lower bit-depth subvolumes: MSBV $\mathbf{X}^M=(X_1^M, X_2^M,\cdots, X_T^M)$ and LSBV $\mathbf{X}^L=(X_1^L, X_2^L,\cdots, X_T^L)$: 
\begin{equation}
    \mathbf{X}^M=\left\lfloor \mathbf{X}/2^{d}\right\rfloor, \ \
    \mathbf{X}^L=\mathbf{X}\  \operatorname{mod} \ 2^{d}, 
\end{equation} 
where $d$ is an integer \textcolor{black}{that represents the cutting position during bit-depth splitting}, satisfying $1\leq d\leq16$. 

The framework of our method is illustrated in Fig.\;\ref{fig: overview}. The whole compression process is divided into two stages. The first stage is utilizing traditional codecs to compress the MSBV for low computational complexity. The second stage is to adopt a slice-by-slice encoding process for each LSBS.
During encoding of the current LSBS $X_{t}^L$, TFAM produces the aligned feature $C_t^a$ utilizing current and preceding MSBS $X_t^M$ and $X_{t-1}^M$, as well as preceding LSBS $X_{t-1}^L$. The TFAM leverages both intra-slice and inter-slice redundancy through cross-attention and self-attention mechanisms. Subsequently, PACM utilizes masked convolution \cite{pixelrnn2016icml} to extract local context $C_t^l$, fused with the aligned feature $C_t^a$, to estimate the probability distribution of $X_{t}^L$. After entropy coding, we concatenate the bitstreams of MSBV and LSBV to obtain the final results. During the decoding process, MSBV is decoded first, which then guides the slice-by-slice decoding process of each LSBS.

\subsection{MSBV Compression}
\label{sec: MSB compression}
As illustrated in Fig.~\ref{fig:visual_image:msb}, employing the bit division approach, the MSBS contains sparse structural information. This sparsity indicates a similarity among pixels in proximate regions, thereby simplifying the compression process.
As reported in Table \ref{table: BPV_results_bits-division}, when using JPEG-XL \cite{alakuijala2019jpeg} to compress the MSBV and LSBV of Covid-CT \cite{morozov2020mosmeddatacovid} and Chaos-CT \cite{kavur2021chaos} datasets, the BPV of MSBV is only $6.37\%$ and $6.54\%$ of LSBV, respectively. This highlights that the compressed MSBV contributes significantly less to the overall bitrates compared to the compressed LSBV. For the sake of low complexity, we directly employ traditional codecs to compress MSBV rather than learning-based methods. 
Considering the limitations inherent in volumetric codecs such as JP3D \cite{schelkens2006jpeg2000p10}, which are constrained in performance, and HEVC \cite{sullivan2012overviewhevc}, known for slower encoding speeds, a balanced approach is vital. In light of this, we adopt JPEG-XL, the cutting-edge traditional image codec, for the compression of MSBV. Additionally, to capitalize on the inter-slice redundancy of MSBV, a method that concatenates all slices in the MSBV vertically is employed. This concatenation increases the area of similar regions in the resultant image, thereby enhancing JPEG-XL’s compression efficiency. The codec's decision tree construction process, particularly its global similarity utilization, benefits significantly from these enlarged areas of similarity.

\begin{table}[!t]
\begin{center}
\caption{Compression performance of MSBV and LSBV in terms of Bits Per Voxel (BPV) using JPEG-LS and JPEG-XL on the Covid-CT and Chaos-CT datasets.}
\label{table: BPV_results_bits-division}
\begin{tabular}{ccccc}
\toprule
\multicolumn{1}{c}{\multirow{2}{*}{Codecs}} & \multicolumn{2}{c}{Covid-CT}  & \multicolumn{2}{c}{Chaos-CT} \\
\cmidrule(lr){2-5}
\multicolumn{1}{c}{} & \multicolumn{1}{c}{MSBV} & \multicolumn{1}{c}{LSBV} & \multicolumn{1}{c}{MSBV} & \multicolumn{1}{c}{LSBV}   \\ 
\midrule

JPEG-LS & \multicolumn{1}{c}{0.488} &  \multicolumn{1}{c}{5.003} & \multicolumn{1}{c}{0.629} & \multicolumn{1}{c}{6.765}  \\

\midrule
JPEG-XL & \multicolumn{1}{c}{0.320} & \multicolumn{1}{c}{5.027}  & \multicolumn{1}{c}{0.434}  & \multicolumn{1}{c}{6.637}\\

\bottomrule
\end{tabular}
\end{center}
\vspace{-0.2cm}
\end{table}

\subsection{LSBV Compression}

As depicted in Figure \ref{fig:visual_image:lsb}, the LSBS encompasses complementary texture information and is depicted with more random noise, presenting more challenges for compression. We introduce learning-based methods to enhance the compression of the LSBV. 
Assuming the real unknown distribution of the LSBV is denoted as $p(\mathbf{X}^L)$, its corresponding Shannon entropy can be expressed as $H(p) = \mathbb{E}_{\mathbf{X}^L \sim p}[-\log(p(\mathbf{X}^L))]$. This represents the theoretical lower bound of the bitrates for compressed images \cite{shannon1948mathematical}. In practical terms, we construct a tractable distribution $p_{\theta}(\mathbf{X}^L)$ that approximates $p(\mathbf{X}^L)$ by minimizing the cross-entropy, which is formulated as:
\begin{equation}
    H(p, p_{\theta}) = \mathbb{E}_{\mathbf{X}^L \sim p}\left[-\log(p_{\theta}(\mathbf{X}^L))\right]. 
\end{equation}
A slice-wise autoregressive model is constructed to leverage the inter-slice redundancy. The distribution of the LSBV can be factorized as $p_{\theta}(\mathbf{X^L}) = \prod_{t=1}^{T} p_{\theta}(X^{L}_{t}|X^{L}_{<t})$. However, for volumetric data with numerous slices, using all previously encoded slice as conditions is impractical. We find that incorporating only the immediately preceding slice significantly reduces inter-slice redundancy while maintaining manageable computational complexity. Consequently, the probability of $p_{\theta}(X^{L}_{t}|X^{L}_{<t})$ is simplified to Markov chain model $p_{\theta}(X^{L}_{t}|X^{L}_{t-1})$. To further exploit intra-slice redundancy, we use $X^M_t$ and $X^M_{t-1}$ as additional conditions, enhancing the estimation of the conditional probability distribution for $X^L_t$.
The probability distribution of $X^L_t$ is thus represented as $p_{\theta}(X^{L}_{t}|X^{M}_{t}, X^{M}_{t-1}, X^{L}_{t-1})$. The corresponding cross-entropy, $H(p, p_{\theta})$, is given by:
\begin{equation}
    H(p, p_{\theta}) = \mathbb{E}_{\mathbf{X}^L \sim p}\left[\sum_{t=1}^{T} -\log p_{\theta}(X^{L}_{t}|X^{M}_{t}, X^{M}_{t-1}, X^{L}_{t-1})\right].
\end{equation}

In this paper, we introduce the Transformer-Based Feature Alignment Module (TFAM), which integrates $X^{M}_{t}$, $X^{M}_{t-1}$, and $X^{L}_{t-1}$ to generate aligned features. Subsequently, the Parallel Autoregressive Coding Module (PACM) fuses aligned features with local context to estimate the probability distribution of the current LSBS, compressing it into a bitstream.

\begin{figure}[!t]
\centering
\includegraphics[width=0.98\linewidth]{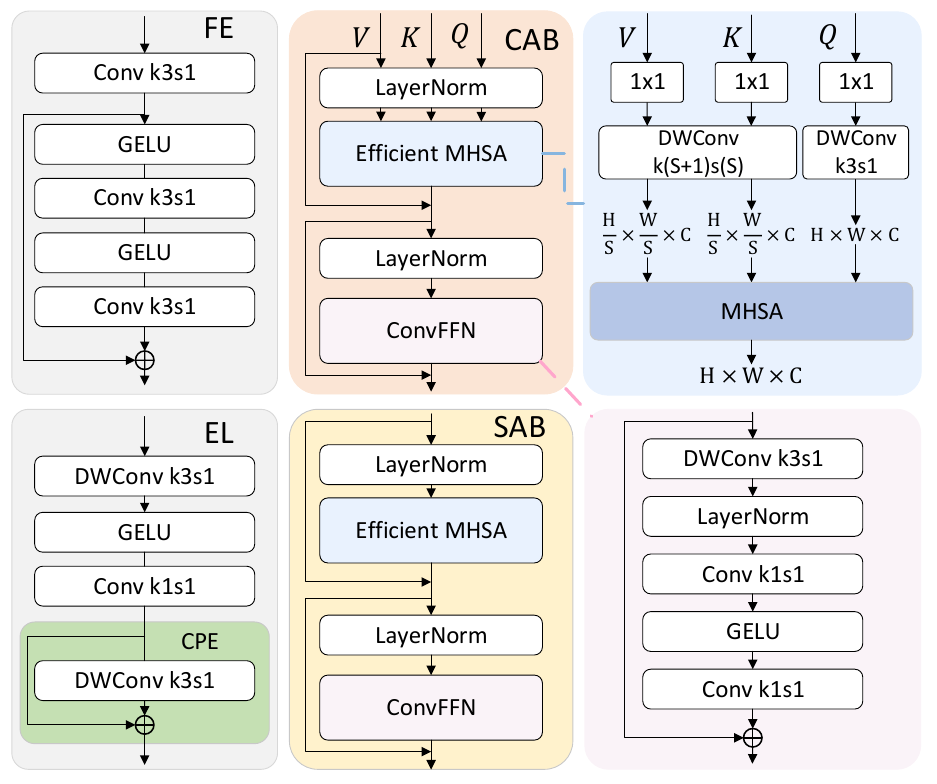}
\caption{
The details of our proposed TFAM, which integrates key elements: Feature Extraction (FE), Embedding Layers (EL), Cross-Attention (CAB), and Self-Attention Blocks (SAB). Notably, $\rm Conv k3s1$ represents a convolution layer with a $3 \times 3$ kernel and stride 1. $\rm 1\times1$ denotes $\rm Conv k1s1$, and $\rm DWConv$ signifies a depth-wise convolution layer. CPE stands for Conditional Position Embedding.
}
\label{fig: TFAM}
\vspace{-0.1cm}
\end{figure}

\subsubsection{Transformer-Based Feature Alignment Module}

We employ a transformer-based model for executing this complex transformation, effectively capturing both short-term and long-term dependency. As shown in Fig. \ref{fig: overview}, the TFAM is utilized to extract features that are aligned with $X^L_t$:
\begin{equation}
C^a_t = {\rm TFAM}(X^{M}_{t}, X^{M}_{t-1}, X^{L}_{t-1}).
\end{equation}
Consequently, the distribution $p_{\theta}(X^{L}_{t}|X^{M}_{t}, X^{M}_{t-1}, X^{L}_{t-1})$ is replaced by $p_{\theta}(X^{L}_{t}|C^a_t)$. The comprehensive structure of TFAM, depicted in Fig. \ref{fig: TFAM}, includes feature extraction layers, embedding layers, and two attention blocks, each consisting of a cross-attention and a self-attention block.

Initially, TFAM processes inputs $X^M_t$, $X^M_{t-1}$, and $X^L_{t-1} \in \mathbb{R}^{H\times W}$ via feature extraction layers. These features undergo further processing in embedding layers and are rearranged to form the query $Q$, key $K$, and value $V$, each in $\mathbb{R}^{n\times c}$, where $n = H \times W$. A Conditional Position Embedding (CPE) \cite{chu2021conditional} is integrated within the embedding layer for enhanced spatial context. These tokens are then processed through two attention blocks to generate the aligned feature. In the first attention block, tokens from $Q$ and $K$ interact with those from $V$ in the cross-attention block, followed by refinement using the self-attention block. Before entering the second attention block, the fused tokens pass through another embedding layer to produce value $V$. New embedding layers regenerate the query $Q$ and key $K$, adapting to the transformed value $V$. The output after the second attention block is the aligned feature $C^a_t$.

Both cross-attention and self-attention blocks consist of an Efficient Multi-Head Self-Attention (EMHAS) layer and a Convolutional Feed-Forward Network (ConvFFN). As the computational and memory costs of the transformer scale quadratically with the resolution of the inputs, to optimize computational efficiency and reduce memory usage, depth-wise convolution with stride $s$ is applied in EMHSA to decrease the spatial dimensions of $K$ and $V$ before attention processing, \emph{i.e.}, $K' = {\rm DWConv}(K)$, $V' = {\rm DWConv}(V) \in \mathbb{R}^{n/{s^2}\times c}$. We set stride $s$ to 4 in our implementation. The efficient attention operation is defined as:
\begin{equation}
{\rm EAttn}(Q,K,V) = {\rm Softmax}\left(\frac{Q{K'}^T}{\sqrt{d}}\right)V'.
\end{equation}
Moreover, diverging from the original Feed-Forward Network (FFN) in the Vision Transformer (ViT) \cite{dosovitskiy2020imagevit}, we replace the linear layer with depth-wise convolution to integrate local information with minimal increase in computational demand:
\begin{equation}
\begin{aligned}
{\rm ConvFFN}(X) &= {\rm Conv}({\rm Conv}(\mathcal{F}(X))) + X, \\
\mathcal{F} &= {\rm LN}({\rm DWConv}(X)),
\end{aligned}
\end{equation}
where the GELU activation function is omitted for simplicity.

\begin{figure}[!t]
\centering
\includegraphics[height=3cm]{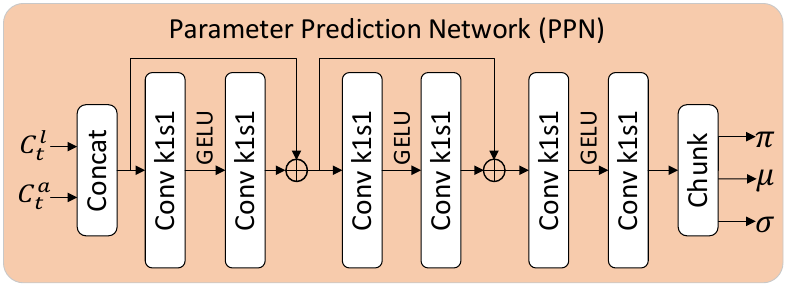}
\caption{The details of parameters prediction networks. The parameters of mixture logistic model $\boldsymbol{\pi}$, $\boldsymbol{\mu}$ and $\boldsymbol{\sigma}$ are generated by PPN.}
\label{fig: PPN}
\vspace{-0.2cm}
\end{figure}

\begin{figure}[!t]
\centering
\includegraphics[height=11cm]{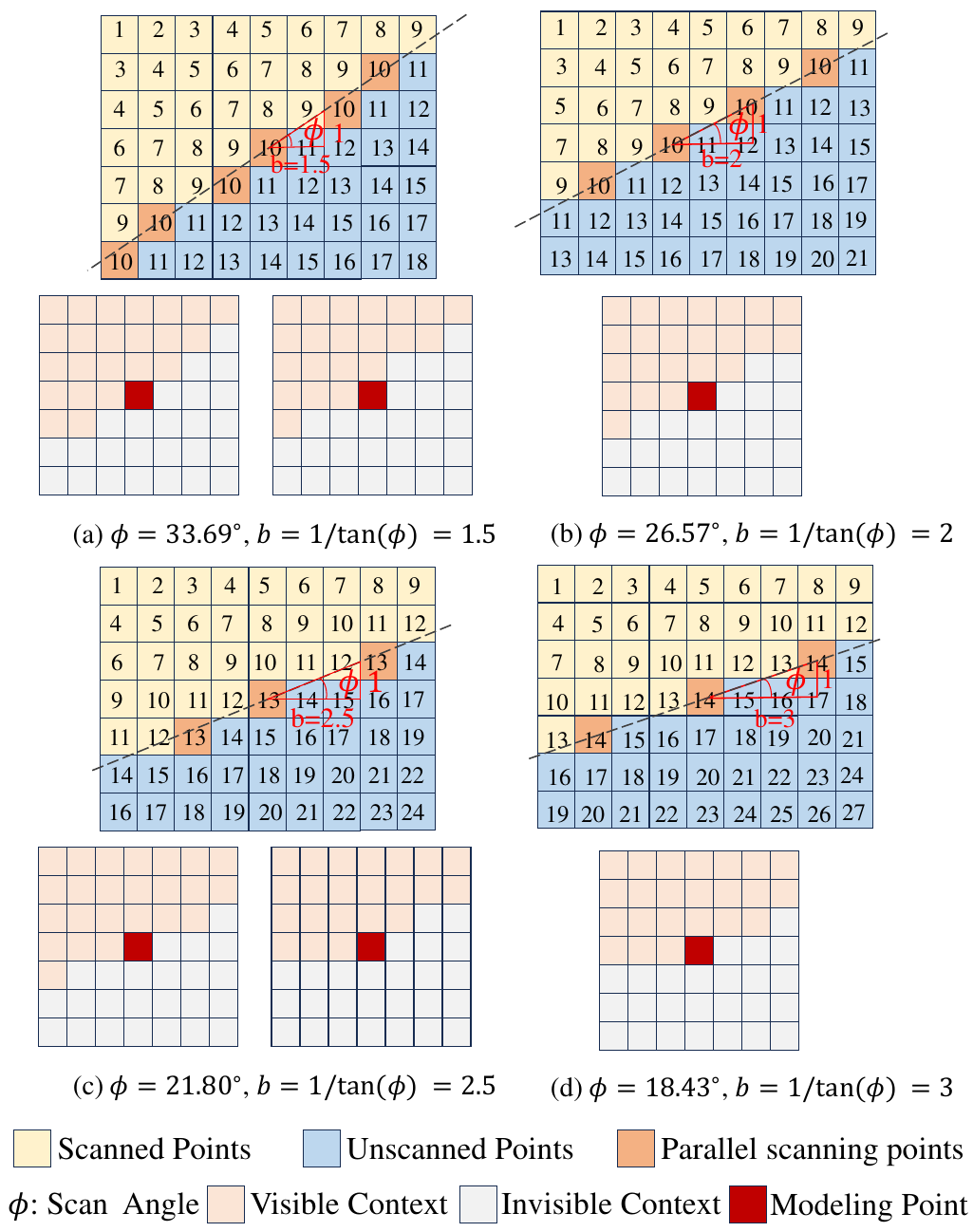}
\caption{\textcolor{black}{Each sub-image's upper half illustrates the relationship between $b$ and different scanning tilt angles, as well as the number of scanning steps for an image of size $7\times 9$ under various scanning sequences. The lower half of each sub-image shows the mask configuration when achieving the maximum receptive field with a $7\times7$ masked convolution at the current scanning angle. When $b$ is an integer, the mask scheme for all pixel points is the same. However, when $b$ is not an integer, it is necessary to design different schemes for pixels in different rows.}}
\label{fig: masked_convolution}
\vspace{-0.1cm}
\end{figure}

\subsubsection{Parallel Autoregressive Coding Module} 
\label{sec: pacm}
Inspired by \cite{zhang2021out,bai2022deep}, we implement a local autoregressive model to utilize the spatial redundancy inherent in the LSBS. This model is conditioned on the aligned feature $C_t^a$ to accurately model the distribution of $X^L_t$. The process of PACM is illustrated in Fig. \ref{fig: overview}. Initially, masked convolution is employed to extract the local spatial context, denoted as $C_t^l$. This context is then concatenated with the aligned feature $C^a_t$ and input into the Parameter Prediction Network (PPN). The PPN generates parameters for the mixture logistic model, which models the distribution of $X^L_t$. Subsequently, $X^L_t$ is encoded into bitstreams using arithmetic coding \cite{arithmetic_coding}.

A single masked convolution layer is utilized to effectively capture the local spatial context, denoted as $C^l_t$, from a specified region. Combining $C^l_t$ with $C^a_t$, the distribution $p_\theta(X^{L}_{t}|X^{M}_{t}, X^{M}_{t-1}, X^{L}_{t-1})$ is replaced by $p_\theta(X^{L}_{t}|C^l_t, C^a_t)$. In the PPN, we fuse $C^l_t$ and $C^a_t$ and, following the approach in \cite{bai2022deep}, model $p_{\theta}(X^{L}_{t}|C^l_t, C^a_t)$ with a logistic mixture model. As shown in Fig. \ref{fig: PPN}, we generate the mixture weights $\boldsymbol{\pi}$, means $\boldsymbol{\mu}$, and variances $\boldsymbol{\sigma}$ for the logistic mixture model. The fusion of $C^l_t$ and $C^a_t$ is achieved through stacking multiple $1 \times 1$ convolution layers. The use of $1\times1$ convolution layers allows the modeling of channel correlations while avoiding information outside the local receptive field. Given $\boldsymbol{\pi}$, $\boldsymbol{\mu}$, and $\boldsymbol{\sigma}$, the probability distribution is given by:
\begin{equation}
    \begin{aligned}
        p_{\theta}(X^{L, ij}_t|C^l_t, C^a_t) = \sum_{n=1}^{N} \pi_n^{(i,j)} {\rm logistic}\left(\mu_n^{(i,j)}, \sigma_n^{(i,j)}\right),
    \end{aligned}
    \label{eq:p_lmm}
\end{equation}
where ${\rm logistic}(\cdot)$ refers to the logistic distribution \textcolor{black}{\cite{pixelcnn_pp}} and $N$ represents the number of mixture components. Unlike color images which possess three channels, medical images typically have only one channel. With \eqref{eq:p_lmm}, the cross-entropy $H(p, p_{\theta})$ can be formulated as:
\begin{equation}
\begin{aligned}
        H(p, p_\theta) = \mathbb{E}_{\mathbf{X}^L\sim p}\left[\sum_{t=1}^{T}- \log p_{\theta}(X^{L}_{t}|C^{l}_t, C^a_t)\right],
\end{aligned}
\label{eq:final_loss}
\end{equation}
which indicates the corresponding compression performance of $X^L$. We employ  \eqref{eq:final_loss} as the loss function in our framework and minimize it through end-to-end training.

\textbf{Parallel Acceleration}: The autoregressive model's requirement for serialized decoding notably hampers coding speed. \textcolor{black}{In the field of lossy compression, spatial or channel autoregression methods \cite{he2021checkerboard,minnen2020channel,li2024groupedmixer} are employed to accelerate the autoregressive process of de-redundant latent features. However, lossless compression operates on pixel domains with lots of redundancy, necessitating the design of pixel-domain-oriented autoregressive acceleration algorithms to ensure the performance of lossless compression \cite{bai2022deep}.} Several strategies are tailored to accelerate the coding speed of the autoregressive model.
Firstly, a block strategy is adopted, where the original slice is divided into $H'\times W'$ patches, which significantly reduces the scanning steps in raster scan order to $H'\times W'$.
\textcolor{black}{Secondly, we introduce a novel formulation to define the relationship between the scanning steps $M$ and the scanning angle $\phi$ for a given patch resolution of $H'\times W'$. We first define $b$ as the weighted coefficient of the $H^{'}-1$ term when calculating the scanning steps, which is determined by the scanning angle $\phi$:
\begin{equation}
\label{eq:b}
    b=\min \left(W^{'}, \frac{1}{\tan{\phi}}\right),
\end{equation}
then the relationship between the number of scanning steps $M_\phi$ and $b$ is as follows:
\begin{equation}
\label{eq:M}
    M_{\phi} = \lceil b(H^{'}-1)1 \rceil + W^{'}.
\end{equation}
As known from Equation (\ref{eq:b}), $b$ remains $b\in [0, W^{'}]$. When $b$ is an integer, its value exactly equals the horizontal distance between parallel scanning points. Moreover, when $b$ is an integer, calculating $M_{\phi}$ does not require rounding, so the maximum visible context range for each pixel is identical. In modeling the context, this allows for the construction of the same mask method for each pixel. When $b$ is a floating-point number, due to the presence of rounding, the maximum visible context range varies across different rows of pixels, necessitating different mask configurations.}

\textcolor{black}{To explore the relationship between the number of masked convolutions and the scanning angle, we first define the rounding method for $b(H'-1)$ as rounding up. The scanning sequence at the start of each row is given by $M_\phi^{(h,1)} = \lceil b(h-1) \rceil + 1$.
We define a difference sequence $O$:
\begin{equation}
    O_i = M_\phi^{(i+1,1)} - M_\phi^{(i,1)}. 
\end{equation}
Suppose the minimal cyclic length of the current difference sequence is $F$, then there are $F$ different context ranges, where the $i$-th row and the $(i+F)$-th row have the same visible context range. When the half-size of the convolutional kernel exceeds the length of the cyclic boundary $F$, i.e., $\lfloor \frac{K}{2} \rfloor \geq F$, the number of required masked convolutions is equal to $F$. Conversely, when the kernel size is smaller, there may be instances where the local context of some rows remains consistent, resulting in a number of convolutional kernels fewer than $F$.}

\textcolor{black}{If $b$ can be expressed as a fraction $b = p/q$, where $p$ and $q$ are coprime, then the denominator $q = F$ represents the length of the cycle. For example, if $b$ is an integer, each difference is consistently the size of $b$, and the cyclic length is $1$, requiring only one type of masked convolutional kernel. If $b = 1.5 = 3/2$, the cyclic length is $2$, with a consistent difference of $3$ every two steps, as shown in Figure~\ref{fig: masked_convolution}(a). In this case, the convolution kernel size $\lfloor \frac{K}{2} \rfloor=3$ exceeds the cyclic length $F=2$, thereby necessitating two types of masked convolutional kernels. Finally, if $b$ cannot be expressed in fractional form, the difference sequence $O$ exhibits no cyclicity, leading to distinct visible context ranges for each row of pixels. In this case, up to $H'$ types of masked convolutions are required to fully encompass the largest receptive field.}

\begin{table*}[!t]
\begin{center}
\vspace{-2mm}
\caption{Summary of the volumetric image datasets used in our experimental studies.}
\label{table: datasets}
\begin{tabular}{cccccccc}
\toprule
Name         & Species          & Position & Type  & Bit Depth & signed & Resolution ($C, H, W$)   &Train/Val Set Size\\ 
\midrule
Heart-MRI    & human            & Heart    &MRI  & 16         & \ding{55} & $90\sim180\times 320 \times 320$    &   20/10     \\
Chaos-CT      & human            & Abdominal &CT  & 16         & \ding{51} & $77\sim105\times512\times512$   &    28/12 \\
Covid-CT     & human            & Lung      &CT  & 16         & \ding{51} & $32\sim72\times512\times512$     & 125/45     \\
Trabit-MRI   & human            & Brain     &MRI  & 16         & \ding{55} & $176\times176\times208$     &  70/30        \\
MRNet    & human            & Knee     &MRI  & 8        & \ding{55} & $17\sim61\times256\times256$     &  1130/120    \\
FAFB-EM      & adult drosophila & Brain  & EM  & 8          & \ding{55} & $64\times64\times64$     & 512/64            \\
\bottomrule
\end{tabular}
\end{center}
\vspace{-0.1cm}
\end{table*}

\section{EXPERIMENT}

\subsection{Datasets}
As shown in Table\;\ref{table: datasets}, we conduct a quantitative evaluation utilizing six distinct volumetric datasets.

\begin{itemize}
\item \textbf{Heart-MRI dataset \cite{tobon2015benchmarkheartmri}} includes $30$ heart MRI volumes. Following \cite{xue2022aiwave},  we crop the images into $64\times64\times64$ blocks and scale them to 16 bits through min-max normalization. The resulting dataset encompasses 283 training blocks and 64 testing blocks.

\item\textbf{Chaos-CT dataset \cite{kavur2021chaos}} contains abdominal CT and MRI volumes from 40 patients. We select CT volumes for compression. Following \cite{xue2022aiwave}, all volumes are cropped into $64\times64\times64$ blocks. The dataset comprises 978 blocks for training and 64 blocks for testing.

\item\textbf{Covid-CT dataset \cite{morozov2020mosmeddatacovid}} comprises pulmonary CT volumetric images. \textcolor{black}{Aligning with the settings of BCM-Net \cite{liu2024bilateral}, we utilize the CT-2 dataset for training and the CT-3 dataset for testing.} Due to the potential presence of metallic objects within the body, the dataset exhibits a value range of [-32146, 32249]. We add 32768 to all data, thereby converting them into positive numbers while minimizing interference with other bit planes.

\item\textbf{Trabit-MRI dataset \cite{mader2019trabit2019}} consists of brain MRI volumes in the Neuroimaging Informatics Technology Initiative (NIfTI) format \cite{larobina2014medical}. This dataset contains a training set, which consists of 70 volumes, and a validation set, encompassing 120 volumes.

\item\textbf{MRNet dataset \cite{bien2018deepmrnet}} is comprised of $1370$ knee MRI exams. Each examination includes three directional scans, referred to as “Axial”, “Coronal” and “Sagittal”. This dataset has been split into a training set consisting of $1130$ exams, and a validation set encompassing 120 exams.

\item\textbf{FAFB-EM dataset \cite{zheng2018completeFAFB}} is an adult fly brain electron microscope anisotropic imaging. In our experiments, we use the setting of aiWave \cite{xue2022aiwave}, which constructs a dataset with $64\times64\times64$ blocks. The training set contains 512 blocks, and the validation set contains 64 blocks.

\end{itemize}

\subsection{Experiment Settings}
\subsubsection{Implementation Details}
For all 16-bit datasets, $d$ is set to 8, except the Heart-MRI dataset. Owing to specific preprocessing steps applied to the Heart-MRI dataset, such as min-max normalization \cite{xue2022aiwave}, we adjust $d$ to 12 to achieve better compression performance.
For 8-bit datasets, $d$ is set to 6 to reduce the randomness of LSBV. In the TFAM architecture, the feature extraction channels are set to 96. Post-embedding features have 192 channels, and during attention, we use 8 heads, each with 32 channels. In the PACM, the masked convolution features have 160 channels, and the mixture components are set $N=10$. The input and output feature channels for the PPN are respectively set as 256 and 30. We also implement our intra-slice method, BD-LVIC-Intra, which compresses every MSBS individually and the input of TFAM is only the current MSBS. Additionally, for our block strategy, we set the patch dimensions to $H' = W' = 32$.

\subsubsection{Training Details}
\label{sec: training details}
LSBS is initially randomly cropped into smaller patches with dimensions of $32 \times 32$. We utilize the Adam optimizer \cite{kingma2015adam} for training, setting the batch size to $32$. 
For standard 16-bit datasets, including Chaos-CT, Covid-CT, and Trabit-MRI, a single model with a consistent set of parameters is trained over 2.2 million iterations. In contrast, for each of the other datasets, separate models are trained with 1.5 million iterations each, thereby developing unique sets of parameters tailored to each dataset.
The initial learning rate is established at $5 \times 10^{-5}$ and is reduced by half at $40\%$, $60\%$, and $80\%$ of the maximum iteration count. During training, we employ data augmentation techniques including random horizontal and vertical flips. Our method is developed using the PyTorch framework. All experiments are conducted on a machine equipped with an Intel CPU i9-10900K and an NVIDIA GeForce RTX 3090 GPU.

\subsubsection{Evaluation Metric}
Our evaluation employs Bits Per Voxel (BPV) as the principal metric. This is quantified as $B(\mathbf{X})/(T \times H \times W)$, wherein $T$ is the total number of slices in volume $\mathbf{X}$, $H$ and $W$ are the corresponding height and width of each slice. Here, $B(\mathbf{X})$ denotes the total size of compressed bitstreams of the volume $\mathbf{X}$.

\subsubsection{Baseline Codecs}
We conduct a comprehensive comparison of our proposed BD-LVIC framework against various 2D image codecs including PNG, JPEG-LS \cite{weinberger2000loco}, JPEG2000 \cite{skodras2001j2k}, the HEVC-RExt intra-coding scheme~\cite{parikh2017highHEVC}, VVC intra-coding scheme \cite{bross2021overviewvvc}, JPEG-XL \cite{alakuijala2019jpeg}, as well as learned lossless image compression methods such as L3C \cite{Mentzer2019cvprL3C} and LC-FDNet \cite{rhee2022lcfdnet}. Additionally, our comparison extends to 3D image codecs, encompassing JP3D \cite{schelkens2006jpeg2000p10}, JPEG2000-Part2 \cite{jp2kp2}, HEVC alongside its range extension strategy (i.e., HEVC-Rext \cite{parikh2017highHEVC}), FFV1 \cite{niedermayer2013ffv1}, VVC \cite{bross2021overviewvvc} and learned lossless volumetric compression methods including ICEC \cite{chen2022exploitingicec} and aiWave \cite{xue2022aiwave}. We reimplement L3C and LC-FDNet to fit 16-bit datasets. For ICEC and aiWave, we report the compression performance published by their authors.

\begin{table*}[t!]
\begin{center}
\vspace{-2mm}
\caption{\textcolor{black}{The Bits Per Voxel (BPV) of 2D and 3D image codecs on 16-bit datasets. Codecs signed by $^*$ indicate learned methods. The most optimal results are indicated in bold, while the second-best outcomes are underlined for clarity.}}
\label{table: BPV_results_16bit}
\begin{tabular}{ccc|cc|cc|cc}
\toprule
\multicolumn{1}{c}{\multirow{3}{*}{Codec}}   &\multicolumn{2}{c|}{{Heart-MRI}}     &\multicolumn{2}{c|}{Chaos-CT}    &\multicolumn{2}{c|}{Covid-CT}      &\multicolumn{2}{c}{Trabit-MRI}    \\ 
\cmidrule(lr){2-9}
&\multirow{2}{1.25cm}{\centering BPV $\downarrow$}     &\multirow{2}{1.35cm}{\centering Compression Ratio  $\uparrow$}   &\multirow{2}{1.25cm}{\centering BPV $\downarrow$}     &\multirow{2}{1.35cm}{\centering Compression Ratio  $\uparrow$} &\multirow{2}{1.25cm}{\centering BPV $\downarrow$}     &\multirow{2}{1.35cm}{\centering Compression Ratio  $\uparrow$} &\multirow{2}{1.25cm}{\centering BPV $\downarrow$}     &\multirow{2}{1.35cm}{\centering Compression Ratio  $\uparrow$}      \\ 
& & & & &  & & &\\
\midrule
\textit{2D Image Codec}     \\ 
PNG  &12.034	&1.330	&9.012	&1.775	&7.09	&2.257	&3.083	&5.190        \\
JPEG-LS\cite{weinberger2000loco}   &11.049  &1.448  &7.383 	&2.167 	&5.019 	&3.188 	&2.224 	&7.194          \\
JPEG2000\cite{skodras2001j2k} &11.436 	&1.399  &6.872 	&2.328 	&5.318 	&3.009 	&2.579 	&6.204 \\
HEVC-RExt-Intra\cite{parikh2017highHEVC} &11.608 	&1.378 	&6.721 	&2.381 	&5.183 	&3.087 	&2.281 	&7.014 \\
JPEG-XL\cite{alakuijala2019jpeg}   &\underline{9.592} 	&\underline{1.668} 	&6.257 	&2.557 	&\underline{4.773} 	&\underline{3.352} 	&\underline{2.102} 	&\underline{7.612} \\
VVC-Intra \cite{bross2021overviewvvc} & - & - & 6.959  & 2.299   & 5.424  & 2.950  & 2.648  & 6.042 \\
L3C$^*$\cite{Mentzer2019cvprL3C}  &10.378 	&1.542 	&6.393 	&2.503 	&5.639 	&2.837 	&2.245 	&7.127  \\ 
LC-FDNet$^*$\cite{rhee2022lcfdnet} &11.124 	&1.438 	&\underline{6.018} 	&\underline{2.659} 	&5.395 	&2.966 	&2.122 	&7.540 \\ 
BD-VILC-Intra$^*$ (Ours)  &\textbf{9.533} 	&\textbf{1.678}  &\textbf{5.630}   &\textbf{2.842}  &\textbf{4.481} 	&\textbf{3.571}   &\textbf{1.964}   &\textbf{8.147} \\ 
\midrule
\textit{3D Image Codec}      \\
JP3D\cite{schelkens2006jpeg2000p10}  &10.221 	&1.565 	&5.624 	&2.845 	&5.272 	&3.035 	&2.324 	&6.885 \\
JPEG2000-Part2\cite{jp2kp2}  &11.133 	&1.437 	&6.566 	&2.437 	&5.309 	&3.014 	&2.543 	&6.292 \\
HEVC-RExt\cite{parikh2017highHEVC}  &10.507 	&1.523 	&5.931 	&2.698 	&5.144 	&3.110 	&2.124 	&7.533 \\ 
FFV1\cite{niedermayer2013ffv1}  &11.045 	&1.449 	&6.427 	&2.489 	&5.093   &3.142  &2.224 	&7.194   \\
VVC \cite{bross2021overviewvvc} & -  & -    & 6.293 & 2.543   & 5.284 & 3.028    & 2.392 & 6.689  \\
TMIV \cite{boyce2021mpegtmiv}   & -  & -    & 6.292 & 2.543   & 5.380 & 2.974    & 2.985 & 5.360  \\
aiWave$^*$\cite{xue2022aiwave}  & \underline{9.161} 	& \underline{1.747} 	& \underline{5.077} 	& \underline{3.151} 	& 4.910 	& 3.259 	&1.910 	&8.377     \\ 
BCM-Net$^*$\cite{liu2024bilateral} &-  &- &- &- &\underline{4.710}  &\underline{3.397}  & \underline{1.910} &\underline{8.377} \\
BD-VILC$^*$ (Ours)  &\textbf{8.842}   &\textbf{1.810} 	&\textbf{5.012}   &\textbf{3.192} 	&\textbf{4.421}   &\textbf{3.619} 	&\textbf{1.856}  &\textbf{8.621} \\ 
\bottomrule
\end{tabular}
\end{center}
\end{table*}

\begin{table*}[t!]
\begin{center}
\caption{\textcolor{black}{The Bits Per Voxel (BPV) of 2D and 3D image codecs on 8-bit datasets. Codecs signed by $^*$ indicate learned methods. The most optimal results are indicated in bold, while the second-best outcomes are underlined for clarity.}}
\label{table: BPV_results_8bit}
\begin{tabular}{ccc|cc|cc|cc}
\toprule
\multicolumn{1}{c}{\multirow{3}{*}{Codec}}   &\multicolumn{2}{c|}{\multirow{1}{*}{Axial}}     &\multicolumn{2}{c|}{Coronal}    &\multicolumn{2}{c|}{Sagittal}      &\multicolumn{2}{c}{FAFB-EM}    \\  
\cmidrule(lr){2-9}
&\multirow{2}{1.25cm}{\centering BPV $\downarrow$}     &\multirow{2}{1.35cm}{\centering Compression Ratio  $\uparrow$}   &\multirow{2}{1.25cm}{\centering BPV $\downarrow$}     &\multirow{2}{1.35cm}{\centering Compression Ratio  $\uparrow$} &\multirow{2}{1.25cm}{\centering BPV $\downarrow$}     &\multirow{2}{1.35cm}{\centering Compression Ratio  $\uparrow$} &\multirow{2}{1.25cm}{\centering BPV $\downarrow$}     &\multirow{2}{1.35cm}{\centering Compression Ratio  $\uparrow$}      \\ 
& & & & &  & & &\\
\midrule
\textit{2D Image Codec}     \\ 
PNG    &5.341 	&1.498 	&4.559 	&1.755 	&5.565 	&1.438 	&6.302 	&1.269  \\
JPEG-LS\cite{weinberger2000loco} &4.898 	&1.633 	&4.058 	&1.971 	&5.224 	&1.531 	&6.297 	&1.270   \\
JPEG2000\cite{skodras2001j2k}  &5.015 	&1.595 	&4.167 	&1.920 	&5.305 	&1.508 	&6.659 	&1.201 \\
HEVC-RExt-Intra\cite{parikh2017highHEVC} &4.975 	&1.608 	&4.140 	&1.932  &5.306  &1.508 	&6.353 	&1.259   \\
JPEG-XL\cite{alakuijala2019jpeg}  &4.724 	&1.693 	&3.887 	&2.058 	&5.083 	&1.574 	&5.972 	&1.340   \\
VVC-Intra \cite{bross2021overviewvvc} & 4.988   &  1.604   & 4.097   & 1.953  & 5.326  & 1.502  & 6.453  & 1.240  \\
L3C$^*$\cite{Mentzer2019cvprL3C}  &5.176 	&1.546 	&4.346 	&1.841 	&5.526 	&1.448 	&6.192 	&1.292 \\ 
LC-FDNet$^*$\cite{rhee2022lcfdnet} &\underline{4.550} 	&\underline{1.758} 	&\underline{3.724} 	&\underline{2.148} 	&\underline{4.917} 	&\underline{1.627}	&\underline{5.719} 	&\underline{1.399} \\ 
BD-VILC-Intra$^*$ (Ours) &\textbf{4.425} 	&\textbf{1.808} 	&\textbf{3.558}  &\textbf{2.248} 	&\textbf{4.769} 	&\textbf{1.678} 	&\textbf{5.701} 	&\textbf{1.403} \\ 
\midrule
\textit{3D Image Codec}      \\
JP3D\cite{schelkens2006jpeg2000p10}  &4.993 	&1.602 	&4.313 	&1.855 	&5.359 	&1.493 	&6.093 	&1.313 \\
JPEG2000-Part2\cite{jp2kp2}  &4.996 	&1.601 	&4.138 	&1.933 	&5.275 	&1.517 	&6.390 	&1.252 \\
HEVC-RExt\cite{parikh2017highHEVC} &4.956 	&1.614 	&4.126 	&1.939 	&5.295 	&1.511 	&6.226 	&1.285 \\
FFV1\cite{niedermayer2013ffv1}    &4.881 	&1.639 	&4.055 	&1.973 	&5.178 	&1.545 	&6.037 	&1.325  \\
VVC \cite{bross2021overviewvvc}   & 4.951	& 1.616 & 4.086	& 1.958 & 5.309	& 1.507 & 6.366 & 1.257 \\
TMIV \cite{boyce2021mpegtmiv}     & 4.989	& 1.604 & 4.088	& 1.957 & 5.309	& 1.507 & 6.367 & 1.256 \\
ICEC$^*$\cite{chen2022exploitingicec}  &4.642 	&1.723 	&3.841 	&2.083 	&4.974 	&1.608 &- &-\\
aiWave$^*$\cite{xue2022aiwave}  &4.545 	& 1.760 	& 3.804 	& 2.103 	& 4.829	& 1.657 	& \underline{5.636} 	& \underline{1.419}   \\ 
BCM-Net$^*$\cite{liu2024bilateral} & \underline{4.410} 	& \underline{1.814} 	& \underline{3.630} 	& \underline{2.204} 	& \underline{4.810} 	& \underline{1.663} & - & -\\  
BD-VILC$^*$ (Ours)  &\textbf{4.322} 	&\textbf{1.851} 	&\textbf{3.461} 	&\textbf{2.311} 	&\textbf{4.679} 	&\textbf{1.710} 	&\textbf{5.413} 	&\textbf{1.478}  \\ 
\bottomrule
\end{tabular}
\end{center}
\vspace{-0.2cm}
\end{table*}

\begin{table}[!t]
\begin{center}
\caption{The average GPU time per slice and number of parameters of our BD-LVIC and other methods on 16-bit Covid-CT \cite{morozov2020mosmeddatacovid} and Trabit-MRI \cite{mader2019trabit2019} datasets (in seconds), with resolutions of $512\times512$ for Covid-CT and $176\times208$ for Trabit-MRI. 'Enc. T.' and 'Dec. T.' represent the encoding and decoding times, respectively.}
\label{table: runtime}
\begin{tabular}{p{1.25cm}ccccc}
\toprule
\multicolumn{1}{c}{\multirow{2}{*}{Codec}} &\multicolumn{1}{c}{\multirow{2}{1cm}{\centering \#Params (M)}} &\multicolumn{2}{c}{Covid-CT} &\multicolumn{2}{c}{Trabit-MRI}       \\ 
\cmidrule(lr){3-4}
\cmidrule(lr){5-6}
& &Enc. T. &Dec. T. &Enc. T. &Dec. T.  \\ \midrule 
\textit{Traditional} \\
\centering JPEG-LS   &-   & 0.068   & 0.067    & 0.030    & 0.029     \\
\centering  JPEG-XL   &-   & 0.254    & 0.120      & 0.065    & 0.044   \\
\centering  HEVC-RExt-Intra &- & 0.649    & 0.024    & 0.064     & 0.002  \\
\centering JP3D      &-   & 0.067   & 0.067     & 0.007     & 0.007    \\
\centering HEVC-RExt &-   & 5.607    & 0.025     & 0.367       & 0.002 \\
\centering VVC       &-   & 54.140    & 0.036     & 3.353       & 0.003 \\
\midrule
\textit{Learned} \\
\centering  L3C   &5M   &8.901        &8.678     &0.411	    &$0.351$       \\
\centering LC-FDNet  & 58.4/56.3M      &2.739          &2.683            &$0.439$	    &$0.433$	       \\
\centering ICEC   & 17.7M   &-          &-            & -	    &-       \\
\centering aiWave   & 695.5M   &-          &-            & -	    &-       \\
\centering BCM-Net   & 12.5M   &-          &-            & -	    &-       \\
\centering Ours    &5.6M   &0.649        &0.648     &0.192	    &0.206	   \\ 
\bottomrule
\end{tabular}
\end{center}
\vspace{-0.1cm}
\end{table}

\subsection{Results}
\subsubsection{Main Compression Results}
In Table\;\ref{table: BPV_results_16bit}, we report the lossless compression results of both 2D and 3D codecs on 16-bit testing datasets. Notably, our proposed BD-LVIC framework outperforms all existing 2D and 3D image codecs. In comparison to the current SOTA method, BCM-Net, our BD-LVIC shows an average BPV performance improvement of $4.5\%$. Against aiWave, the average enhancement is $4.4\%$, and BD-LVIC surpasses the leading traditional codec, JPEG-XL, by $11.7\%$.
Concurrently, our intra-slice compression scheme, BD-LVIC-Intra, exceeds all 2D image codecs, and in some datasets, even surpasses aiWave. This demonstrates the scalable ability of our framework to apply for 2D medical images. On different datasets, there is a noticeable gain disparity between BD-LVIC-Intra and BD-LVIC. This variation underscores the impact of inter-slice redundancy on enhancing 3D codec performance, which is largely influenced by slice thickness. For instance, the thicknesses of the Chaos-CT and Covid-CT datasets are $3mm$ and $8mm$, respectively. Thinner slices typically indicate higher inter-slice similarity, thereby amplifying the performance gap between 2D and 3D codecs. 
Furthermore, in Table\;\ref{table: BPV_results_8bit}, we detail the lossless compression results on 8-bit testing datasets. BD-LVIC also exhibits superior results in compressing 8-bit medical datasets. The quantified results on both 16-bit and 8-bit datasets demonstrate that our BD-LVIC framework achieves state-of-the-art lossless compression performance and general applicability for volumetric medical images.

\subsubsection{Coding Speed}
Due to the unavailability of ICEC, aiWave, and BCM-Net in open-source, we retrained two rapid learning-based image methods for a comparative analysis of encoding and decoding times against our framework. Table\;\ref{table: runtime} presents the encoding and decoding performance of our framework. Despite the integration of a transformer-based and an autoregressive module, the BD-LVIC framework consistently achieves remarkable encoding and decoding speeds. Specifically, On the Covid-CT dataset, our encoding time is recorded at 0.649 seconds, which is on the same order of magnitude as JPEG-XL and significantly outperforms HEVC-RExt, L3C, and LC-FDNet. 
We further evaluate the coding speed on 8-bit datasets. On the FAFB dataset, our framework achieves an encoding time per slice of only 0.13 seconds, significantly faster than aiWave's \cite{xue2022aiwave} 943.211 seconds. On the Coronal subset of MRNet \cite{bien2018deepmrnet}, BCM-Net's encoding time per volume is 179.299 seconds, markedly slower than our framework's 6.333 seconds. Furthermore, our framework's parameter count is comparable to that of L3C, enhancing its suitability for deployment on edge devices.

\subsubsection{Bitrates Allocation}
\label{sec: bitrates allocation}
As reported in Table\;\ref{table: percentage_bitrates}, we present the bitrates of both the MSBV and the LSBV, along with their respective proportions of the total bitrates. Across five datasets, LSBV's bitrates forms a range of $88.09\%$ to $94.48\%$ of the total, averaging at about $92.63\%$. This indicates that the LSBV exhibits a larger scope for optimization. Designing more complex compression modules for the LSBV could yield considerable additional compression gains.

\subsection{Ablation Study}
In our ablation study, to streamline the process, we do not train the BD-LVIC-Intra framework to compress the first slice, because our primary objective is evaluating the impacts of both intra-slice and inter-slice references on compression efficacy. We adapt BD-LVIC by training it with unique parameters for each dataset, deviating from the unified parameter used for Chaos-CT and Covid-CT datasets as detailed in Section\;\ref{sec: training details}. Consequently, the compression results of the ablation study yield marginally superior compression performance compared with Table\;\ref{table: BPV_results_16bit} and Table\;\ref{table: percentage_bitrates}.

\begin{table}[!t]
\begin{center}
\caption{Compression performance of BD-LVIC framework: evaluating MSBV, LSBV, and total bitrates across various datasets measured by BPV. The $(\cdot)$ denotes the percentage contribution to the total bitrates.}
\label{table: percentage_bitrates}
\begin{tabular}{cccc}
\toprule
\multicolumn{1}{c}{\multirow{1}{*}{Dataset}} &\multicolumn{1}{c}{MSBV} &\multicolumn{1}{c}{LSBV} &\multicolumn{1}{c}{Total}  \\ \midrule

Heart-MRI & 0.488 ($5.52\%$) & 8.354 ($94.48\%$) & 8.842 \\
Chaos-CT &0.287 ($5.73\%$) & 4.725 ($94.27\%$) & 5.012 \\
Covid-CT &0.294 ($6.65\%$) & 4.127 ($93.35\%$) & 4.421 \\
Trabit-MRI & 0.221 ($11.91\%$) & 1.635 ($88.09\%$) & 1.856 \\
Coronal & 0.243 ($7.02\%$) & 3.218 ($92.98\%$) & 3.461 \\
\bottomrule
\end{tabular}
\end{center}
\vspace{-0.1cm}
\end{table}

\subsubsection{Analysis of TFAM} 
In Table\;\ref{table: align_by_TFAM}, we evaluate the efficacy of employing our proposed TFAM for aligning adjacent slices to augment the compression performance of LSBV. Compared with the approach of utilizing resblock \cite{he2016deep} and concatenation operations for merging the context from the current MSBS, preceding MSBS and LSBS, our TFAM, which generates aligned context, achieves BPV enhancements of 0.294 and 0.189 on the Heart-MRI and Chaos-CT datasets, respectively.

\begin{table}[!t]
\begin{center}
\vspace{-1mm}
\caption{Compression performance of LSBV for Covid-CT and Heart-MRI with and without TFAM.}
\label{table: align_by_TFAM}
\begin{tabular}{ccccc}
\toprule
Method &\multicolumn{1}{c}{Heart-MRI} &\multicolumn{1}{c}{Chaos-CT}        \\ \midrule
Aligned by TFAM  & 8.330 & 4.712 \\
w/o TFAM   &8.624 (+0.294)  &4.901 (+0.189) \\
\bottomrule
\end{tabular}
\end{center}
\end{table}

\begin{table}[!t]
\centering
\vspace{-2mm}
\caption{The Bits per voxel (BPV) of our BD-LVIC on the LSBV under different combinations of the $X_{t-1}^L$, $X_t^M$, and $X_{t-1}^M$.}
\label{table: results_different_inputs}
\begin{tabular}{cccccc}
\toprule
$X_{t-1}^L$ & $X_{t}^M$ & $X_{t-1}^M$ & Heart-MRI & Chaos-CT & Covid-CT \\
\midrule
\ding{51} & \ding{51} & \ding{51} & 8.330 & 4.712 & 4.086 \\
\ding{55}  & \ding{51} & \ding{51}    & 8.821 & 5.180  & 4.116 \\
\ding{55}  & \ding{55}    & \ding{51} & 9.242 & 5.398 & 4.407 \\
\ding{55}   & \ding{55}    & \ding{55}    & 9.594 & 5.820  & 4.507 \\
\bottomrule
\end{tabular}
\end{table}

\subsubsection{Analysis of Bit Division}
\begin{figure}[!t]
\centering
\includegraphics[width=0.9\linewidth]{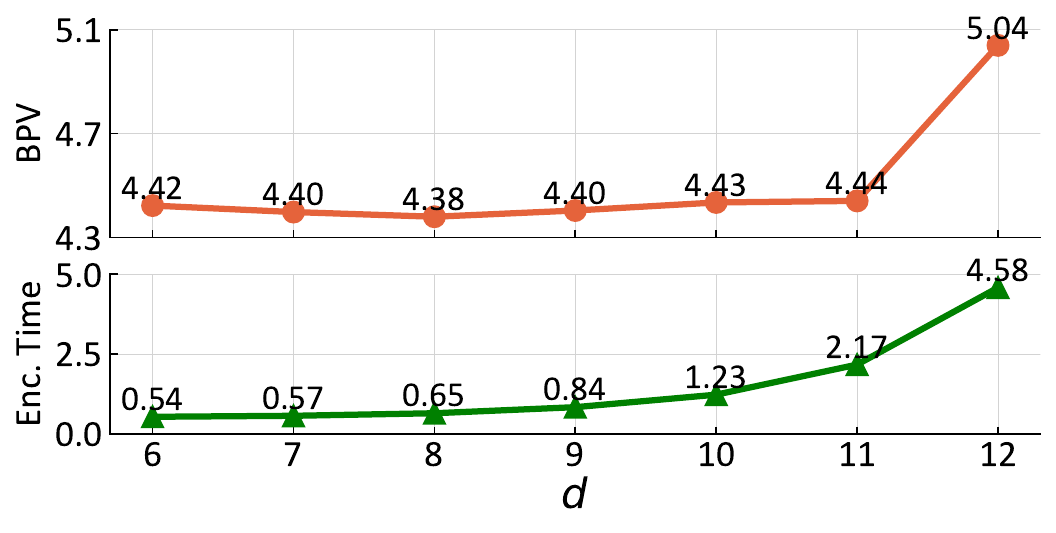}
\setlength{\abovecaptionskip}{-1pt}
\caption{Evaluating BD-LVIC framework's performance on Covid-CT dataset: impact of bit division positions (denoted as $d$) on BPV and per-slice encoding time (Enc. Time) on GPU, measured in Seconds.}
\label{fig: different_d}
\end{figure}

As illustrated in Fig.\;\ref{fig: different_d}, we assess the impact of bit division position, $d$, on the compression efficacy of our BD-LVIC framework. We observe a trend where the BPV first decreases and then increases as $d$ ranges from $6$ to $12$. 
This can be explained by the increased randomness in the distribution of the LSBV at lower bit division positions, which amplifies the challenge of compression. In contrast, higher bit division positions broaden the data range, complicating the probability distribution estimation.
At $d=12$, there's a marked increase in overall BPV. Beyond the expansion of the data range, this increase is partly attributed to Covid-CT images typically being padded with negative values like $-2048$. When $d<12$, these values become 0 in the LSBV, but at $d=12$, the corresponding values in LSBV transform into 2048, challenging the adaptability of a single autoregressive model to such non-zero large padding values. Meanwhile, with the increasing bit depth of LSBV, the construction time for the PMF table increases rapidly, resulting in a substantial rise in the overall encoding latency.

\textcolor{black}{Additionally, we evaluated the effectiveness of our joint compression scheme under extreme conditions, specifically when $d=12$. We compared two compression schemes at $d=12$ on the Covid-CT dataset: our joint compression scheme and the separate compression of the MSBV and LSBV. The BPV for our joint compression scheme and the separate compression scheme are 5.04 and 5.227, respectively. The joint compression scheme demonstrates a 3.58\% improvement in performance compared to the separate scheme. These results affirm the robustness of our joint compression framework in handling challenging scenarios.}

\begin{figure}[!t]
\centering
\includegraphics[width=0.9\linewidth]{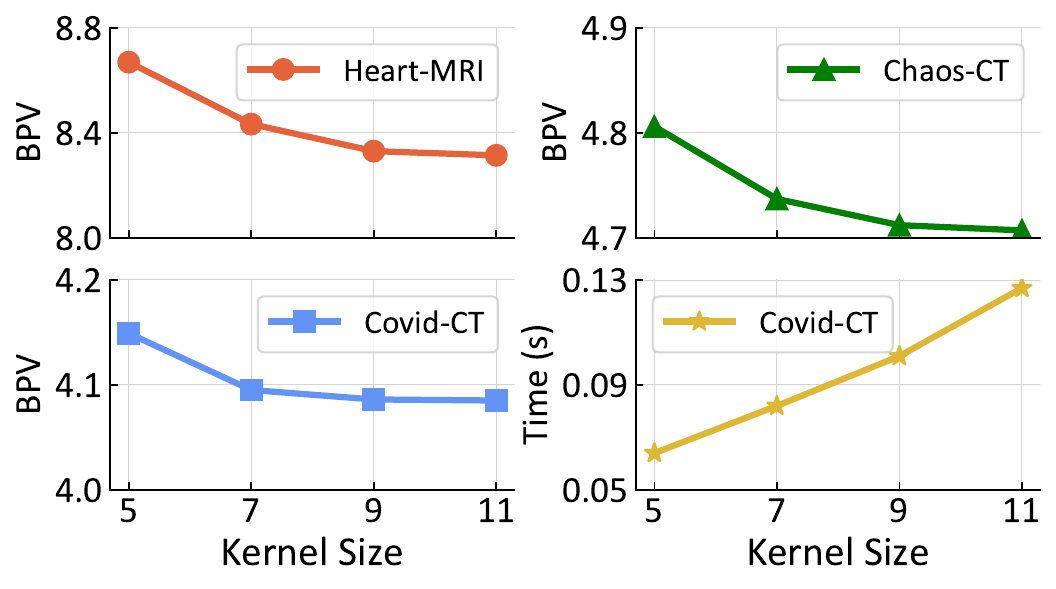}
\caption{Comparative assessment of compression efficiency of LSBV in diverse datasets employing masked convolution layer with different kernel sizes. 'Time' denotes the per-slice execution duration of masked convolution on the Covid-CT dataset, measured in seconds.}
\label{fig: different_kernel_size}
\vspace{-1mm}
\end{figure}

\subsubsection{Analysis of Intra-slice and Inter-slice Dependency}
As reported in Table\;\ref{table: results_different_inputs}, our ablation study investigates the impact of incorporating information from $X_t^{M}$, $X_{t-1}^{M}$ and $X_{t-1}^L$. By changing the inputs of TFAM, we can analyze the effectiveness of intra- and inter-slice dependency on the compression performance of LSBV. In datasets like Heart-MRI and Chaos-CT, which exhibit higher inter-slice similarity, all three inputs significantly enhance the compression performance. Conversely, in datasets such as Covid-CT, characterized by lower inter-slice similarity, the $X_{t}^M$ plays a predominant role in boosting compression efficacy.

\subsubsection{Analysis of Kernel Size of Masked Convolution}
As discussed in Section\;\ref{sec: pacm}, the selection of scan angles directly influences the quantity of available context. With an established scan angle, larger convolution kernel sizes can be employed to capture more context, thereby enhancing compression performance. As illustrated in Fig.\;\ref{fig: different_kernel_size}, we examine the impact of varying convolutional kernel sizes on the compression performance and inference time of LSBV. An increase in the kernel size from 5 to 9 leads to notable performance gains. However, the transition from 9 to 11 yields only marginal benefits. Considering the additional almost $50\%$ increase in parameters and $26\%$ increase in inference time from sizes 9 to 11, we ultimately opt for a kernel size of 9.

\begin{figure}[!t]
    \centering
    \includegraphics[width=0.85\linewidth]{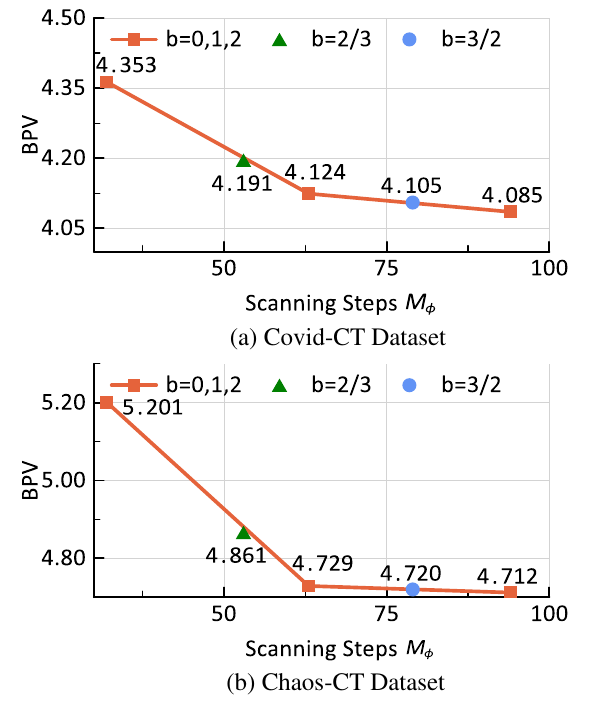}
    \caption{\textcolor{black}{Variations in compression performance (BPV) of LSBV at different scanning steps, where (a) and (b) represent results on the Covid-CT\cite{morozov2020mosmeddatacovid} and Chaos-CT \cite{kavur2021chaos} datasets, respectively. The slice thickness for the Covid-CT dataset is 8mm, while that for the Chaos-CT dataset is 3mm.}}
    \label{fig:bpv_scannning_step}
    \vspace{-1mm}
\end{figure}

\subsubsection{Analysis of Scanning Angle}
\textcolor{black}{We evaluated the compression performance of the proposed parallel scan order versus the traditional raster scan order. On the Covid-CT dataset, the BPV for the parallel scan order (b=2) and raster scan order are 4.38 and 4.37, respectively, indicating comparable compression efficiency. Notably, the parallel scan order requires only 94 steps, compared to $H^{'} \times W^{'} = 1024$ steps for the raster scan, significantly reducing scanning steps.
}

\textcolor{black}{Then, we further tested the model's compression performance and scanning steps under different scanning angles with convolutional kernel size $K = 9$. The test results are shown in Figure~\ref{fig:bpv_scannning_step}. The Figure~\ref{fig:bpv_scannning_step} aims to elucidate two main points. First, we think of a piecewise linear association between the number of scanning steps and the compression efficacy, characterized by varying slopes across discrete integer intervals. For instance, the intervals from $b=0$ to $b=1$ and from $b=1$ to $b=2$ demonstrate a roughly linear relationship, albeit with differing slopes. This observation is supported by the data points at $b=2/3$ and $b=3/2$, which lie on the linear regressions of their respective integer intervals.}

\begin{table}[!t]
\centering
\caption{\textcolor{black}{The correlation between scanning steps $M_{\phi}$ and encoding time for the Chaos-CT dataset \cite{kavur2021chaos}.}}
\begin{tabular}{cccc}
 \toprule
 $b$ & $\phi$  & $M_\phi$  &   Encoding Time (s)\\
 \midrule
 0   & $90^{\circ}$       &    32            &  0.054  \\
 2/3 & $63.43^{\circ}$    &    53            &  0.080  \\
 1   & $45^{\circ}$       &    63            &  0.092  \\
 3/2 & $33.69^{\circ}$    &    79            &  0.113  \\
 2   & $26.57^{\circ}$    &    94            &  0.131  \\
 \bottomrule
\end{tabular}
\label{table:scanning_steps_encoding_time}
\end{table}

\textcolor{black}{Second, the optimal balance between performance and scanning steps, characterized by the scanning angle $\phi$, varies across datasets, primarily due to differences in the slice thickness of medical images. In datasets with thinner slices, such as the Chaos-CT dataset (slice thickness 3mm), adjacent slices provide better inter-slice references for the current modeling point, thereby reducing the dependence on local information within the slice. Consequently, even larger scanning angles can still achieve excellent compression performance. Conversely, in datasets with thicker slices, such as Covid-CT (slice thickness 8mm), there is a greater reliance on local information within the slice, necessitating smaller scanning angles to ensure effective compression performance. Our proposed 2D line scanning algorithm allows for flexible adjustment of scanning angles from $0^{\circ}$ to $90^{\circ}$, enabling a more precise approximation of the optimal balance between inference speed and compression performance, tailored to different requirements for bandwidth and inference speed in various application scenarios. Our primary consideration was compression performance; therefore, we ultimately selected the scanning sequence $b = \frac{1}{\tan(26.57^\circ)} = 2$.}

\textcolor{black}{Additionally, as shown in Table \ref{table:scanning_steps_encoding_time}, we tested the impact of scanning steps on compression speed on the Chaos-CT dataset. The results in the table indicate that reducing the number of scanning steps can significantly accelerate the compression speed of the model. Due to the inherent parallel acceleration capabilities of cuDNN, the scanning steps on both the encoding and decoding sides are identical, ensuring consistency between encoding and decoding processes.}

\section{Conclusion}
In this paper, we presented BD-LVIC framework for high bit-depth volumetric medical images, effectively addressing the degraded performance, excessive memory usage, and low throughput of existing learning-based methods. BD-LVIC decomposes a high bit-depth volume into two lower bit-depth subvolumes: MSBV and LSBV. For the MSBV, we employ traditional codecs to reduce computational complexity. Additionally, we introduce the TFAM and PACM, enabling efficient compression of the LSBV.
Extensive experiments demonstrate that the BD-LVIC framework achieves not only state-of-the-art lossless compression performance but also competitive coding speed for volumetric medical image compression tasks.
\bibliographystyle{ieee}
\bibliography{ref.bib}

\end{document}